\documentclass[12pt]{article}
\usepackage{amsmath}
\renewcommand{\a}{\alpha} 
\renewcommand{\b}{\beta}
\renewcommand{\d}{\delta} 

\newcommand{\g}{\gamma}

\renewcommand{\l}{\lambda} 
 
\newcommand{\m}{\mu} 
\newcommand{\n}{\nu}

\newcommand{\s}{\sigma}
\renewcommand{\S}{\Sigma} 

\renewcommand{\t}{\tau}

\newcommand{\tV}{\tilde{V}}

\newcommand{\cL}{{\cal L}}

\newcommand{\pd}{\partial}

\newcommand{\half}{\frac{1}{2}}

\begin{document}
\pagestyle{myheadings}
\markboth{\today}{}

\author{Jiri Hoogland\footnote{jiri@cwi.nl} 
  and Dimitri Neumann\footnote{neumann@cwi.nl}\\
  CWI, P.O.~Box 94079, 1090 GB  Amsterdam, The Netherlands
  }
\title{\textbf{Scaling invariance in finance II:\\
    Path-dependent contingent claims}}
\maketitle

\thispagestyle{empty}
\begin{abstract}
  This article is the second one in a series on the use of scaling
  invariance in finance. In the first
  article~\cite{HooglandNeumann99a}, we introduced a new formalism for
  the pricing of derivative securities, which focusses on tradable
  objects only, and which completely avoids the use of martingale
  techniques. In this article we show the use of the formalism in the
  context of path-dependent options. We derive compact and intuitive
  formulae for the prices of a whole range of well known options such
  as arithmetic and geometric average options, barriers, rebates and
  lookback options. Some of these have not appeared in the literature
  before. For example, we find rather elegant formulae for double
  barrier options with moving barriers, continuous dividends and all
  possible configurations of the barriers. The strength of the
  formalism reveals itself in the ease with which these prices can be
  derived. This allowed us to pinpoint some mistakes regarding
  geometric mean options, which frequently appear in the literature.
  Furthermore, symmetries such as put-call transformations appear in a
  natural way within the framework.
\end{abstract}

\newpage

\section{Introduction}
\label{sec:introduction}

In our previous paper~\cite{HooglandNeumann99a} we introduced a new
formalism for the pricing of derivative securities, based on the idea
of `relativity' of prices.  The core of this formalism is the idea
that problems should be formulated in terms of {\it tradable objects
  only}. (Note that we use a broad definition of the term tradable:
every quantity that can be represented by a self-financing portfolio
is considered to be a tradable). If this is done, we can show that
functions expressing the price of a derivative in terms of prices of
the underlying tradables should always be homogeneous of degree one.
This follows from a simple dimensional analysis argument. The same
should therefore be true for payoff functions, in terms of which the
contracts are specified. In fact, we claim that {\it any payoff
  function should be representable by a homogeneous function of degree
  one in tradables}, for else it is {\it ill-defined}.  In this paper
we show that the formalism works very well for path-dependent options,
and that this leads to more compact formulae, which can be verified
more easily and have a clear financial interpretation.

Our formulation is based on a PDE-approach. We have shown that, by
using homogeneity properties, it is very easy to derive a PDE
describing the evolution of claim prices, without making use of any
martingale techniques. The solution of the PDE may, of course, be cast
in the form of a Feynman-Kac formula using a Green's function approach.
The PDE has an explicit symmetry corresponding to the freedom of
choice of a numeraire.  The PDE is defined in terms of volatility
functions of tradables only, drift terms are irrelevant. The only
place where drift terms do play a role is in the analysis of
arbitrage: if deterministic relations exist between tradables, there
are conditions on the drift terms in order to exclude arbitrage.  The
fact that our PDE has an explicit numeraire independence is
fundamental. It is in marked contrast with the usual Black-Scholes PDE
approach. In the latter approach the numeraire, some currency, is
fixed in advance and bonds nominated in this currency are considered
to be risk-free. This, of course, is an illusion. Even if interest
rates are constant, a dollar bond is not risk-free in the eyes of a
European investor. In fact, the only object that truly deserves the
name `risk-free' is an object that has zero-value at all times. To
declare any other object to be risk-free breaks the symmetry and makes
calculations very much less transparent. An important lesson that
modern physics teaches us is that symmetries which exist in a problem
should be preserved in every step of a calculation.  It makes the
formalism transparent and gives us a powerful tool to verify results.
We want to put forward the opinion that {\it scaling symmetry is one
  of the most fundamental ingredients for a pricing theory}, and that
other concepts, like the existence of an equivalent martingale
measure, should be derived from this, not the other way round.

Finally, it should be mentioned that many tricks and symmetries that
appear in an ad hoc way in the literature, for example put-call
symmetries and similarity reductions, can be traced back to the
fundamental property of homogeneity of pricing functions. It places
all these concepts in a unified framework.

The outline of this article is as follows.  In
Section~\ref{sec:homogeneity} we recall some of the results from
Ref.~\cite{HooglandNeumann99a}. We show how to derive the fundamental
PDE using the homogeneity properties of pricing functions, given a set
of tradables with stochastic dynamics driven by Brownian motion. Next,
we give an algorithm which is used to derive claim prices using this
PDE. Then we show that the PDE posses a symmetry, associated with
numeraire independence, which implies that only the {\it relative}
volatilities of the underlying tradables with respect to each other
matter.  Furthermore, it is shown that the homogeneity leads in a very
straightforward way to generalized put-call symmetries. Finally, we
recall the general solution for the case of European claims in a
lognormal world, using Green's functions.  In Section~\ref{sec:asians}
we show that the formulation of the asian claim pricing problem in
terms of the tradables leads to a considerable simplification of the
governing PDE. We show that some results regarding geometric average
options which appear frequently in the literature
(e.g.~\cite{Wilmott98}) contain mistakes.  We provide the correct
solutions.  In Section~\ref{sec:more-solutions} we construct a large
class of solutions to the PDE, new tradables, which turn out to be
very useful in the pricing of barrier-type options in a very compact
and intuitive way. In Section~\ref{sec:barriers-and-friends} we
discuss the pricing of single and double barrier options, rebates and
lookback options. We derive very clean formulae and show the various
symmetries between them. We correct and extend some results on double
barrier options. We end with a discussion and outlook.

\section{Homogeneity and contingent claim pricing}
\label{sec:homogeneity}

In the first paper~\cite{HooglandNeumann99a} in this series we have
shown that a fundamental, but so far apparently overlooked, property
of any properly defined market of tradables is that the price of any
claim depending on other tradables in the market should be a
homogeneous\footnote{A function $f(x_0,\ldots,x_n)$ is called
  homogeneous of degree $r$ if $f(ax_0,\ldots,ax_n)=a^r
  f(x_0,\ldots,x_n)$.  Homogeneous functions of degree $r$ satisfy the
  following property (Euler): $\sum_{\m=0}^nx_\m\frac{\pd}{\pd
  x_\m}f(x_0,\ldots,x_n)=r f(x_0,\ldots,x_n)$} function of degree one
of these same tradables. This property is nothing but a consequence
of the simple fact that prices of tradables are only defined
with respect to each other. Let us assume that we have a market of
$n+1$ basic tradables with prices $x_\mu$ ($\mu=0,\ldots,n$) at time
$t$. The price of any tradable in this market with a payoff depending
on the prices of these basic tradables should satisfy the following
scaling symmetry:
\[
V(\l x,t)=\l V(x,t)
\]
which automatically implies (Euler)
\[
V(x,t)=x_\m \pd_{x_\m}V(x,t)
\]
where $\pd/\pd x_\m\equiv\pd_{x_\m}$\footnote{We will always use Greek
  symbols for indices running from $0$ to $n$ and Latin symbols for
  indices running from $1$ to $n$. Furthermore, we use Einstein's
  summation convention: repeated indices in products are summed over.}
. We use this fundamental property to derive a general PDE, giving the
price of such a claim in a world where the dynamics of the tradables
are driven by $k$ independent standard Brownian motions, as
follows\footnote{The dot denotes an inner-product w.r.t. the $k$
  driving diffusions.}
\[
dx_\m=x_\m(\s_\m(x,t)\cdot dW(t)+\a_\m(x,t) dt)
\]
where we assume $\s_\m$ and $\a_\m$ are properly chosen and no
summation over indices is assumed. Note that we do not specify the
numeraire in terms of which the drift and volatility are expressed. It
does not matter. Applying It\^o to $V(x,t)$ we get
\[
dV(x,t)=\pd_{x_\m}V(x,t)dx_\m+\cL V(x,t)dt
\]
where 
\[
\cL V(x,t)\equiv 
\bigg(
\pd_t+\half\s_\m(x,t)\cdot\s_\n(x,t)x_\m x_\n\pd_{x_\m}\pd_{x_\n}
\bigg)V
\]
So, if $V(x,t)$ solves $\cL V=0$ with the payoff at maturity as
boundary condition $V(x,T)=f(x)$, then we immediately have a
replicating self-financing trading strategy because of the homogeneity
property.  We will drop the distinction between such derived and basic
quantities and always refer to them as tradables. Note that we do not
have to use any change of measure to arrive at this result. The drifts
are irrelevant for the derivation of the claim price. Only the
requirement of uniqueness of the solution, i.e. no arbitrage, leads to
constraints on the drifts terms if deterministic relations exist
between the various tradables~\cite{HooglandNeumann99a}.

\subsection{The algorithm}
\label{sec:algorithm}

To price contingent claims we start out with a basic set of tradables.
Using these tradables we may construct new, derived, tradables, whose
price-process $V$ depends upon the basic tradables.  Of course, these
new tradables should be solutions to the basic PDE, $\cL V=0$. Their
payoff functions serve as boundary conditions (Note that prices of
basic tradables trivially satisfy the PDE, by construction). If the
derived tradables are constructed in this way, we can use them just
like any other tradable. In particular, we can use them as underlying
tradables, in terms of which the price of yet other derivative claims
can be expressed (and so on...)  In fact, this is a fundamental
property that any correctly defined market should posses, and we will
come back to this point in Ref.~\cite{HooglandNeumann99c}. It amounts
to a proper choice of coordinates to describe the economy.

\vspace{1\baselineskip} \noindent The general approach to the pricing
of a (European) path-dependent claim in our formalism can be described
as follows.
\begin{enumerate}
\item The payoff is written in terms of tradable objects.  (At this
  point, it might already be necessary to introduce new, derived
  tradables. This is illustrated in Section~\ref{sec:asians} on
  Asians).
\item A PDE is derived for the claim price with respect to these
  tradables.
\item The PDE is solved.
\item Consistency check: the solution should be independent of the
  numeraire.
\end{enumerate}

\subsection{Symmetries of the PDE}
\label{sec:symmetries-pde}
The scale invariance of the claim price is inherited by the PDE via an
invariance of the solutions of the PDE under a simultaneous shift of
all volatility-functions by an arbitrary function $\l(x,t)$
\[
\s_\m(x,t)\to\s_\m(x,t)-\l(x,t)
\]
This corresponds to the freedom of choice of a numeraire. It just
states that volatility is a relative concept. Price functions should
not depend on the choice of a numeraire. This can easily be checked by
noting that for homogeneous functions of degree $1$ we have
\[
x_\m \pd_{x_\m}\pd_{x_\n}V=0
\]
This ensures that terms involving the $\l$ drop out of the PDE. Note
that this equation gives an interesting relation between the various
$\Gamma$'s of the claim. This symmetry is in marked contrast with the
standard Black-Scholes equation, in which a gauge has been fixed.

\subsection{Generalized put-call symmetries}
\label{sec:generalized-put-call}

As an example of the strength of this symmetry, and to show the
natural embedding in our formalism, consider an economy with two
tradables with prices denoted by $x_{1,2}$ and dynamics given by
($i=1,2$)
\[
dx_i=\s_i(x_1,x_2,t)x_i \cdot dW+\ldots
\]
It is easy to see that under certain conditions there should be a
generalized put-call symmetry. Any claim with payoff $f(x_1,x_2)$ at
maturity and price $V(x_1,x_2,t)$ should satisfy
\[
\bigg(
\pd_t+\half |\s(x_1,x_2,t)|^2 x_1^2 \pd_{x_1}^2
\bigg)V=0
\]
where $\s(x_1,x_2,t)=\s_1(x_1,x_2,t)-\s_2(x_1,x_2,t)$.  Homogeneity
implies that it also solves
\[
\bigg(
\pd_t+\half |\s(x_1,x_2,t)|^2 x_2^2 \pd_{x_2}^2
\bigg)V=0
\]
Therefore, if $|\s(x_1,x_2,t)|^2=|\s(x_2,x_1,t)|^2$, this PDE can be
rewritten as
\[
\bigg(
\pd_t+\half |\s(x_2,x_1,t)|^2 x_2^2 \pd_{x_2}^2
\bigg)V=0
\]
and we see that $V(x_2,x_1,t)$ with payoff $f(x_2,x_1)$ is a solution,
too. This is nothing but a generalized put-call symmetry.  In the
first case $x_2$ acts as numeraire, in the second case $x_1$ takes
over this role. The usual put-call symmetry follows if we take a
constant $\s$ and let $x_1, x_2$ represent a stock and a bond
respectively. This result holds also when early-exercise features are
included, but extra care should be taken in that case to make sure
that the boundary conditions satisfy the symmetry-operation. See for
example Ref.~\cite{CarrChesney96}. Furthermore, we would like to point
out that in the usual formulation, which uses martingales, these
symmetries also appear, but only after a lot of work.

\subsection{Lognormal asset prices}
\label{sec:logn-asset-pric}
In an economy with lognormal distributed asset-prices
\[
dx_\m(t)=\s_\m(t)x_\m(t)\cdot dW(t)+\ldots
\]
it is possible to write down a very elegant formula for European-type
claims, as was shown in Ref.~\cite{HooglandNeumann99a}
\begin{equation}
  \label{eq:1}
  V(x_0,\ldots,x_n,t)=\int V(x_0 \phi(z-\theta_0),\ldots,
  x_n \phi(z-\theta_n),T)d^mz
\end{equation}
with
\[
\phi(z)=\frac{1}{\left(\sqrt{2\pi}\right)^m}
\exp\left( -\frac{1}{2} \sum_{i=1}^{m} z^2_i \right)
\]
The $\theta_\m$ are $m$-dimensional vectors, which follow from a
singular value decomposition of the covariance matrix $\S_{\m\n}$ of
rank $m\le k$:
\[
\S_{\m\n}\equiv\int_t^T \s_{\m}(u)\cdot\s_{\n}(u)du=\theta_\m\cdot\theta_\n
\]

\section{Asians}
\label{sec:asians}

Asian contingent claims provide payoffs which involve average prices
of one (or more) of the underlying tradables. The averaging makes them
strongly path-dependent~\cite{Taleb96}, as knowledge of the price path
is required for the determination of the average. In this section we
discuss arithmetic and geometric average options~\cite{KemnaVorst90}.
We show that by working with tradables it is possible to write down
compact PDE's for the price process. For geometric average strike
options it is shown that certain results, which appear frequently in
the literature (e.g.~\cite{Wilmott98}), contain mistakes.

\subsection{Arithmetic average options}

A general rule is that payoff functions can be expressed as
homogeneous functions of degree $1$ in tradables.  How do we handle
information from the past in this context? Let us consider an
elementary example, a contract that pays $S(s)$ at time $T>s$. A
problem now becomes apparent. How do we relate the value of an object
at two different times? A dollar today is not a dollar tomorrow. So we
have to define how to translate value through time. What usually is
done is to express the value of the stock at time $s$ in a reference
currency (say dollars). Since this is a ratio, it is well-defined.
This is then converted back into dollars at time $T$. In a formula
\[
S(s) \rightarrow \frac{S(s)P(T,T)}{P(s,s)}
\]
Here $P(t,T)$ is a bond which pays $1$ dollar at time $T$.  The next
question is: what tradable has this value at time $T$?  We have to
distinguish two time ranges, $t>s$ where we know the ratio $\d\equiv
S(s)/P(s,s)$ and $t<s$ where we do not.  In the former case, the
tradable is simply $\d P(t,T)$.  In the latter case we can consider it
as an option whose value at time $s$ equals $\d P(s,T)$. Assuming a
lognormal world, it has value
\[
\int \frac{S(t)\phi(z-\theta_S)P(t,T)\phi(z-\theta_{P_T})}
{P(t,s)\phi(z-\theta_{P_s})} dz
\]
Here we used Eq.~\ref{eq:1}.  Therefore, the tradable that should
appear in the payoff is given by
\[
V_{S(s)}=
\left\{ 
  \begin{array}{ll}
    \d P(t,T) & s<t \\
    \frac{S(t)P(t,T)}{P(t,s)}
    \exp(\theta_S-\theta_{P_s})\cdot(\theta_{P_T}-\theta_{P_s})
    & s>t 
  \end{array} 
\right.
\]
This tradable is an elementary building block for many path dependent
options. For example, an arithmetic average price call option, sampled
at discrete times $t_i$ ($1\leq i\leq n$), has a payoff defined by
\[
\left(\frac{1}{n}\sum_{i=1}^n V_{S(t_i)}-K P_T\right)^+
\]
Note that this formulation allows for stochastic interest rates. To
hedge the claim, we need not only the stock $S$ and the bond
$P_T\equiv P(t,T)$ maturing at time $T$, but also as many bonds as
there are sample dates. Obviously, when we consider a continuously
sampled average, this becomes problematic.  So let us consider a
simplification, assuming a deterministic relation between the bond
prices
\[
P(t,s)=e^{r(T-s)}P(t,T)
\]
We will call this a deterministic bond structure. It corresponds to
the assumption that interest rates are constant. The bonds do not have
to be deterministic, the deterministic relations only imply that all
bonds have the same volatility.  In this case the currency $N$ in
which the bond is nominated can be expressed as
\[
N(t) = P(t,t) = e^{r(T-t)} P(t,T)
\]
This corresponds to the usual assumptions in the literature when the
bonds are deterministic.  Now $V_{S(s)}$ reduces to
\[
V_{S(s)}=
\left\{ \begin{array}{ll}
\d P(t,T) & s<t \\
e^{-r(T-s)}S(t) & s>t
\end{array} \right.
\]
Now let us consider options involving a continuously sampled
arithmetic average price. This price is represented by the tradable
\[
\begin{aligned}
\bar{S}(t)&=\frac{1}{T}\int_0^T V_{S(s)}ds \\
          &=   \frac{1-e^{-r(T-t)}}{rT} S(t)
           +\frac{1}{T}\int_0^t \frac{S(s)}{P(s,s)} ds\; P(t,T) \\
          &\equiv\phi S+AP
\end{aligned}
\]
If we take the bond as numeraire, then
\[
dS=\s S dW+\cdots, \hspace{5mm}
d\bar{S}=\phi\s S dW+\cdots
\]
If we choose the bond as numeraire, in this \textsl{gauge} the
relative volatility of the bonds is zero.  This immediately leads to
the following PDE for the price of an arithmetic average option
\[
\bigg(
\pd_t+\frac{1}{2}\s^2S^2(\pd_S+\phi\pd_{\bar{S}})^2
\bigg)V=0
\]
If we perform a change of variables, using the running average $A$
instead of $\bar{S}$, we find
\[
\bigg(
\pd_t+\frac{S}{N(t)T}\pd_A+\frac{1}{2}\s^2S^2\pd_S^2
\bigg)V=0
\]
which corresponds to the usual formulation. Unfortunately it is hard
to find explicit solutions for this equation.  For an example see
\cite{GemanYor93}.

\subsection{Geometric average options}

What about geometric average options? For these, we need another
building block, a tradable paying at time $T$
\[
\log\left(\frac{S(s)}{P(s,s)}\right)P(T,T)
\]
Again we distinguish two time ranges. For $t>s$, the tradable is $\d
P(t,T)$ where $\d\equiv\log(S(s)/P(s,s))$ is known.  For $t<s$ its
value is given by
\[
\int \log\left(\frac{S(t)\phi(z-\theta_S)}
{P(t,s)\phi(z-\theta_{P_s})}\right)P(t,T)\phi(z-\theta_{P_T}) dz
\]
This integral can be evaluated easily, giving
\[
P(t,T)\left(\log\left(\frac{S(t)}{P(t,s)}\right)
-\frac{1}{2}|\theta_S-\theta_{P_T}|^2
+\frac{1}{2}|\theta_{P_s}-\theta_{P_T}|^2\right)
\]
In the deterministic bond setting, things simplify to
\[
V_{\log(S(s))}=
\left\{ \begin{array}{ll}
\d P(t,T) & s<t \\
\left(
  \log\left(\frac{S(t)}{P(t,T)}\right)
  -r(T-s)-\frac{1}{2}\s^2(s-t)
\right)P(t,T) & s>t
\end{array} \right.
\]
(Note: for $s=T$ this is a log-option).  From this we can construct a
tradable representing the logarithm of the geometric average (up to a
factor $T^{-1}$)
\begin{eqnarray*}
  \bar{s}(t)
  &=& \int_0^T V_{\log(S(s))}ds \\
  &=& P(t,T)\left((T-t)\log\left(\frac{S(t)}{P(t,T)}\right)
    -\frac{1}{2}(r+\frac{1}{2}\s^2)(T-t)^2\right) \\
  &&+P(t,T)\int_0^t \log\left(\frac{S(s)}{P(s,s)}\right)ds
\end{eqnarray*}
If we take the bond as numeraire, then
\[
dS=\s S dW+\cdots
,\quad
d\bar{s}=\s (T-t)P dW+\cdots
\]
Therefore, we find the following PDE for claim prices
\[
\bigg(
\pd_t+\frac{1}{2}\s^2(S^2\pd^2_S
+2(T-t)SP\pd_S\pd_{\bar{s}}+(T-t)^2P^2
\pd^2_{\bar{s}})
\bigg)V=0
\]
Now it is useful to perform a change of variables.  We want to trade
in $\bar{s}$ for the following object, which is again a tradable, and
equals the geometric average on expiry
\begin{eqnarray}
  \lefteqn{\tilde{S}=P\exp\left(\frac{\bar{s}}{P T}
      +\frac{\s^2(T-t)^3}{6T^2}\right)} \label{eq:2} \\
  &=&P\exp\left(\frac{1}{T}\left((T-t)\log\left(\frac{S}{P}\right)
      -\frac{1}{2}(r+\frac{1}{2}\s^2)(T-t)^2+I\right)
    +\frac{\s^2(T-t)^3}{6T^2}\right) \notag
\end{eqnarray}
where we have introduced $I$ for the integral over $\log(S/P)$. In
terms of this tradable, the PDE becomes
\[
\bigg(
\pd_t+\frac{1}{2}\s^2\bigg(S^2\pd^2_S
+2\frac{T-t}{T}S\tilde{S}\pd_S\pd_{\tilde{S}}
+\frac{(T-t)^2}{T^2}{\tilde{S}}^2
\pd^2_{\tilde{S}}\bigg)
\bigg)V=0
\]
It turns out that the prices we obtain for geometric average options
differ from those in the literature. What seems to be missing there is
the last term in the exponential in Eq.~\ref{eq:2}. We are convinced
that our results are correct.  To prove this, let us explicitly show
that $\tilde{S}$ satisfies the fundamental PDE
\[
{\cal L}\tilde{S}(S,P,t) 
= \frac{\tilde{S}}{T}\left(
  \pd_t I-\log\left(\frac{S}{P e^{r(T-t)}}\right)
\right)
\]
This vanishes by definition of $I$. Note that the delta's
corresponding to $\tilde{S}$ are simple
\[
\pd_S\tilde{S}=
\left(\frac{T-t}{T}\right)\frac{\tilde{S}}{S}
,\quad
\pd_P\tilde{S}=
\left(\frac{t}{T}\right)\frac{\tilde{S}}{P}
\]
Consider an average price call. Since its payoff $(\tilde{S}-KP)^+$
only depends on $\tilde{S}$ and $P$, the relevant PDE reduces to
\[
\bigg(
\pd_t+\frac{1}{2}\s^2\frac{(T-t)^2}{T^2}{\tilde{S}}^2\pd^2_{\tilde{S}}
\bigg)V=0
\]
The solution of this problem is
\[
V_{GAPC}(S,P,K,t) = \tilde{S}\Phi(d_1)-K P \Phi(d_2)
\]
with
\[
d_{1,2}\equiv\frac{\log\left(\frac{\tilde{S}}{K P}\right)
\pm\frac{1}{2}\S^2}{\S}
,\quad
\S^2\equiv\frac{\s^2(T-t)^3}{3T^2}
\]
Here $\Phi(z)\equiv\int^z_{-\infty}\exp(-\half x^2)dx/\sqrt{2\pi}$. If
we look at the price of the option at $t=0$, and use the currency
corresponding to the bond as numeraire (so that $P=e^{-r(T-t)}$), this
formula reduces to
\begin{eqnarray*}
  \lefteqn{V_{GAPC}(S,P,K,0)=}\\
  &=&e^{-a T}S \Phi\left(
    \frac{\log\left(\frac{S}{K}\right)+aT}{\S}
  \right)
  -e^{-r T}K\Phi\left(
    \frac{\log\left(\frac{S}{K}\right)+ a T-\S^2}{\S}
  \right)
\end{eqnarray*}
Here $a=\frac{1}{2}(r+\frac{\s^2}{6})$. We have written this in the
format used in~\cite{Wilmott98}, but find a different result.
Next consider an average strike call. After a suitable change of
numeraire, the PDE becomes
\[
\bigg(
\pd_t+\frac{1}{2}\s^2\frac{t^2}{T^2}S^2\pd^2_S
\bigg)V=0
\]
In this case, the solution is
\[
V_{GASC}(S,\tilde{S},t) = S\Phi(d_1)-\tilde{S}\Phi(d_2)
\]
with
\[
d_{1,2}\equiv\frac{\log\left(\frac{S}{\tilde{S}}\right)
\pm\frac{1}{2}\S^2}{\S}
,\quad
\S^2\equiv\frac{\s^2(T^3-t^3)}{3T^2}
\]
Again, if we look at the price of the option at $t=0$, and use the
currency corresponding to the bond as numeraire this formula reduces
to
\[
V_{GASC}(S,\tilde{S},0)=S \Phi\left(\frac{aT+\half\S^2}{\S}\right)
-e^{-a T} S \Phi\left(\frac{aT-\half\S^2}{\S}\right)
\]

\section{Generating new solutions}
\label{sec:more-solutions}

In this section we consider claims depending on one lognormal stock
$S$ with volatility $\s$ in a deterministic bond structure as before.
It should be clear that in our approach it is a trivial exercise to
write down the corresponding formula, when we have two lognormal
stocks instead of one stock and one bond. We show that the symmetries
of the PDE allow for the construction of classes of solutions, which
prove very useful in constructing solutions for barrier-type claims.
The governing PDE for the claim-price $V(S,P,t)$ with payoff $f(S,P)$
at time $T$ can be written as
\[
\cL V\equiv \bigg(\pd_t+\half\s^2S^2\pd_S^2\bigg)V=0
\]
In the following we will use time-to-maturity $\t\equiv T-t$ instead
of the running time $t$ to simplify the notation. The PDE becomes
\[
\cL V\equiv \bigg(-\pd_\t+\half\s^2S^2\pd_S^2\bigg)V=0
\]
The Green's function for this PDE is given by
\[
G(x,\t)
=\frac{1}{\s\sqrt{\t}}
\phi\bigg(\frac{\log(x)-\half\s^2\t}{\s\sqrt{\t}}\bigg)
,\quad
\phi(z)\equiv\frac{1}{\sqrt{2\pi}}e^{-\half z^2}
\]
Now we construct a tradable $R_\a(S,P,\t)$ with payoff
$F_\a(S,P)\equiv(S/P)^\a P$ at maturity for some constant $\a$.  The
convolution of the payoff function with the Green's function yields a
new set of tradables
\[
R_\a(S,P,\t)\equiv \left(\frac{S}{e^{-\half\s^2\a\t}P}\right)^\a 
e^{-\half\s^2\a\t}P
\]
Note that $R_0(S,P,\t)=P$ and $R_1(S,P,\t)=S$. These tradables satisfy
two very nice symmetry properties. On the one hand we have
\[
R_\a(S,P,\t)=R_{-\a+1}(P,S,\t)
\]
On the other hand
\begin{eqnarray*} 
  \label{eq:3} 
  R_\a(R_\a(S,P,\t),R_{\a+1}(S,P,\t),\t) &=& S \\
  R_{\a+1}(R_\a(S,P,\t),R_{\a+1}(S,P,\t),\t) &=& P   
\end{eqnarray*}
These are special cases of the following relations
\begin{eqnarray*}
  R_\a(R_\b(S,P,\t),R_{\b+1}(S,P,\t),\t) &=& R_{\b-\a+1}(S,P,\t) \\
  R_\a(R_{\b+1}(S,P,\t),R_\b(S,P,\t),\t) &=& R_{\b+\a}(S,P,\t)
\end{eqnarray*}
If we apply It\^o to $R_\a(S,P,\t)$ we obtain
\[
dR_\a=\a \s R_\a dW+\ldots
\]
Therefore the price $V(\l R_\a,\m R_\b,\t)$ of a claim with payoff
$f(\l R_\a,\m R_\b)$ at maturity with $\l,\m\in{\bf R}$ satisfies the
following PDE:
\[
\bigg(
-\pd_\t+\half(\a-\b)^2\s^2R_\a^2\pd_{R_\a}^2
\bigg)V=0
\]
If we introduce a scaled time $\tilde{\t}\equiv (\a-\b)^2\t$ this PDE
becomes of the form
\[
\bigg(
-\pd_{\tilde{\t}}+\half\s^2R_\a^2\pd_{R_\a}^2
\bigg)V=0
\]
which bears close resemblance to the PDE at the beginning of this
section.  This implies that if $V(S,P,\t)$ solves $\cL V=0$ with the
payoff $V(S,P,0)=f(S,P)$  as boundary-condition, then
\[
V\big(\l R_\a,\m R_\b,(\a-\b)^2\t\big)
\]
is a solution too of the PDE with payoff $f(\l R_\a,\m R_\b,0)$. Let
us consider a few simple examples.  If we denote by $V_C(S,P,\t)$ the
price of a vanilla call with strike $1$ and payoff $f_C(S,P)=(S-P)^+$
then also 
\[
V_C(S,KP,\t)=V_C(R_1,KR_0,\t)
\]
with strike $K$ will be a solution with payoff $f_C(S,KP)$. In a
similar way we define the price $V_P(S,P,\t)$ of a vanilla put with
strike $1$ and payoff $f_P(S,P)=(P-S)^+$.  Automatically
$V_P(S,KP,\t)$ will be a solution, too, with payoff $f_P(S,KP)$. The
put-call parity transformation follows immediately:
\[
V_C(S,KP,\t)=V_P(KP,S,\t)
\]
It also follows trivially that when $K_CK_P=(S/P)^2$,
\[
V_C(S,K_CP,\t)=\sqrt{K_C/K_P}V_P(S,K_PP,\t)
\]
Note that $S/P$ is what is called the forward price in the literature.

Finally, note that we may construct additional solutions to a PDE with
a time-dependent volatility-function $\s(\t)$. In this case the
solutions are of the form $V(\l R_\a,\m R_{\a\pm1},\t)$ where
\[
R_\a(S,P,\t)=\left(\frac{S}{e^{\half\S^2\a}P}\right)^\a e^{\half\S^2\a}P
\]
and $\S^2=\int_0^{T-t}\s(u)^2du$.

The earlier quoted symmetries can be used to relate the various
solutions of the PDE and can be used very fruitfully in the
construction of solutions of, for example, barrier-type options. These
symmetries lead to generalized put-call symmetries, although they also
provide symmetries between vanilla and barrier options, as we will
show in the next section.

These types of symmetries may be very useful when one needs to hedge
an exotic contingent claim, as was already observed in
Ref.\cite{CarrChou97}.  Especially with barrier-type options it allows
one to introduce semi-static hedges.

\section{Barriers and friends}
\label{sec:barriers-and-friends}

In this section we discuss contingent claims which possess so called
weak path-dependence~\cite{Taleb96}. These claims have payoff-features
which are triggered by some event during the life-time of the
contract.  In contrast to the asian-type options, which are strongly
path-dependent, their valuation is less involved, as it does not
require knowledge of the complete path. We first consider single
moving barriers, where we show the usefulness of the symmetry of
solutions discussed in the previous section. Then we move on to double
moving-barriers claims, which can be seen as a double infinite sum of
single moving-barriers claims. Only a few terms are required to obtain
accurate results.  We show that the results of
Ref.~\cite{KunitomoIkeda92} are valid only when $L<K<U$, where $L,U$
denote the lower- and upper-barrier and $K$ the strike, and give the
correct results for the general case including continuous dividends.
Finally, we discuss lookback options and show that they fit nicely
into our framework.

Note that barriers are often monitored at discrete times.  In
Ref.~\cite{BroadieGlassermanKou99} a simple and straightforward way to
correct for this fact has been put forward. It only involves a simple
shift of the barriers depending on the frequency of monitoring.

\subsection{Single barriers}
\label{sec:barriers}

We start our discussion with the simplest type of a single barrier
claim, a down-and-out call. This is a call option with the additional
feature that it becomes worthless when the stockprice hits a barrier,
given by $S=Be^{\g\t}P\equiv B(\t)P$, from above during the lifetime
of the option. Here we have $B,\g\in{\bf R}$. As before, we assume
that we have a deterministic bond structure. This implies that $\g=r$
corresponds to a constant barrier (in terms of the currency in which
the bond is nominated), while other values of $\g$ lead to
exponentially moving barriers. The claim price should satisfy the
standard PDE
\[
\big(-\pd_\t+\half\s^2S^2\pd_S^2\big) V_{DOC}(S,P,K,B,\g,\t)=0
\]
It is well known that all specific properties of this option are in
the boundary-conditions. The boundary-condition for the European
down-and-out call are simply
\begin{alignat*}{3}
  &V_{DOC}(S,P,K,B,\g,0)
  = (S-KP)^+
  &\quad\quad\mathrm{\mbox{for all }} S>BP \\
  &V_{DOC}(B(\t)P,P,K,B,\g,\t)
  = 0 
  &\quad\mathrm{\mbox{for all }} \t>0
\end{alignat*}
We will now assume that $B<K$, and come back to the case $B>K$ later.
Let us first consider the case where we do not have the second
boundary condition and the payoff is defined on the whole positive
$S$-axis. This is just the standard European call $V_C(S,KP,\t)$. The
second boundary condition says that the value of the option becomes
zero if, during its lifetime, the barrier is hit. This implies that
the option price should be lower than that of a plain call since we
take more risk. Thus we have to subtract a barrier-premium from the
standard call price.

Let us write $\tV(S,P,K,B,\g,\t)$ for this barrier premium.  What
boundary conditions should this function obey? Of course, its value
should coincide with $V_C(S,KP,\t)$ on the boundary $S=B(\t)P$ for all
$\t>0$:
\[
V_C(B(\t)P,KP,\t)=\tV(B(\t)P,P,K,B,\g,\t)
\]
If, in addition, we have $\tV(S,P,K,B,\g,0)=0$ for $S>BP$, then we can
construct the solution to the down-and-out call by
\begin{equation}
  \label{eq:4}
  V_{DOC}(S,P,K,B,\g,\t)=V_C(S,KP,\t)-\tV(S,P,K,B,\g,\t)
\end{equation}
At this point we invoke the results of the previous section. If
$V_C(S,P,\t)$ solves $\cL V=0$, then so does
\[
\tV(S,P,K,B,\g,\t) 
= V_C\big(\l R_\a(S,BP,\t),\m R_\b(S,BP,\t),(\a-\b)^2\t\big)
\]
With this choice of the barrier-premium Eq.~\ref{eq:4} satisfies all
boundary conditions if we set
\[
\l=1
,\quad
\m=\frac{K}{B}
,\quad
\a=-\frac{2\g}{\s^2}
,\quad
\b=\a+1
\]
The solution is thus given by
\begin{eqnarray*}
\lefteqn{V_{DOC}(S,P,K,B,\g,\t)=}\\
&=& V_C(S,KP,\t)-V_C\bigg(R_\a(S,BP,\t),\frac{K}{B}R_{\a+1}(S,BP,\t),\t\bigg)
\end{eqnarray*}
If $B>K$, we can apply the same construction. Only the call $V_C$
should now be replaced by a `left-clipped' call $V_C^{B+}$, which can
be defined by its payoff
\[
V_C^{B+}(S,KP,0)=\left\{\begin{array}{ll}
S-KP & \mbox{for $S>BP$}\\
0 & \mbox{otherwise} \end{array} \right.
\]
Thus its value is given by
\[
V_C^{B+}(S,KP,\t)=S\Phi(d_1)-KP\Phi(d_2)
\]
with
\[
d_{1,2}=\frac{\log(\frac{S}{BP})\pm\half\s^2\t}{\s\sqrt{\t}}
\]
In a similar way, we can value an up-and-out call. Such an option is
only interesting for $B>K$, since otherwise it is worthless.
Repeating the same arguments, we see that it can be defined in terms
of a `right-clipped' call $V_C^{B-}$, with payoff
\[
V_C^{B-}(S,KP,0)=\left\{\begin{array}{ll}
S-KP & \mbox{for $KP<S<BP$}\\
0 & \mbox{otherwise} \end{array} \right.
\]
Its value follows from $V_C^{B-}=V_C-V_C^{B+}$.  So the value of the
up-and-out call is given by
\begin{eqnarray*}
\lefteqn{V_{UOC}(S,P,K,B,\g,\t)=}\\
&=&V_C^{B-}(S,KP,\t)-V_C^{B-}
\bigg(R_\a(S,BP,\t),\frac{K}{B}R_{\a+1}(S,BP,\t),\t\bigg)
\end{eqnarray*}
The values of up/down-and-in call follow from in-out parity
\begin{eqnarray*}
  V_{DIC}(S,P,K,B,\g,\t)
  &=&V_C(S,KP,\t)-V_{DOC}(S,P,K,B,\g,\t) \\
  V_{UIC}(S,P,K,B,\g,\t)
  &=&V_C(S,KP,\t)-V_{UOC}(S,P,K,B,\g,\t)
\end{eqnarray*}
Note that under the transformations $R_\a\leftrightarrow S,
R_{\a+1}\leftrightarrow P$ we find a vanilla-barrier transformation
which is valid for $B<K$
\[
V_{DIC}\bigg(R_\a(S,BP,\t),\frac{R_{\a+1}(S,BP,\t)}{B},K,B,\g,\t\bigg)=
V_C(S,KP,\t)
\]
For a down-and-out call we find a similar symmetry, which is
actually valid for all $B$:
\[
V_{DOC}\bigg(R_\a(S,BP,\t),\frac{R_{\a+1}(S,BP,\t)}{B},K,B,\g,\t\bigg)
=-V_{DOC}(S,P,K,B,\g,\t)
\]
And similarly for the up-and-out call. Furthermore, we can immediately
write down the price for all single barrier put options by using the
generalized put-call transformation
\begin{eqnarray*}
  V_{UOP}(S,P,K,B,\g,\t) 
  &=& K V_{DOC}\bigg(P,S,\frac{1}{K},\frac{1}{B},-\g,\t\bigg) \\
  V_{UIP}(S,P,K,B,\g,\t) 
  &=& K V_{DIC}\bigg(P,S,\frac{1}{K},\frac{1}{B},-\g,\t\bigg) \\
  V_{DOP}(S,P,K,B,\g,\t) 
  &=& K V_{UOC}\bigg(P,S,\frac{1}{K},\frac{1}{B},-\g,\t\bigg) \\
  V_{DIP}(S,P,K,B,\g,\t) 
  &=& K V_{UIC}\bigg(P,S,\frac{1}{K},\frac{1}{B},-\g,\t\bigg)
\end{eqnarray*}
It is a simple check to see that these claim prices satisfy all
appropriate boundary conditions.

Let us show that the results above can be rewritten in the more
well-known form using
\begin{eqnarray*}
  R_\a(S,BP,\t) 
  &=& \left(\frac{S}{e^{-\half\s^2\a\t}BP}\right)^\a e^{-\half\s^2\a\t}BP\\
  &=& \left(\frac{S}{e^{\g\t}BP}\right)^\a e^{\g\t}BP
  = \left(\frac{S}{B(\t)P}\right)^\a B(\t)P\\
  R_{\a+1}(S,BP,\t) 
  &=& \left(\frac{S}{e^{-\half\s^2(\a+1)\t}BP}\right)^{\a+1}
  e^{-\half\s^2(\a+1)\t}BP\\
  &=& \left(\frac{S}{e^{\g\t}BP}\right)^{\a}\frac{S}{e^{\g\t}}
  = \left(\frac{S}{B(\t)P}\right)^{\a}\frac{BS}{B(\t)}
\end{eqnarray*}
For example
\begin{eqnarray*}
  \tV(S,P,K,B,\g,\t)
  &=&V_C\bigg(R_\a(S,BP,\t),\frac{K}{B}R_{\a+1}(S,BP,\t),\t\bigg)\\
  &=& \bigg(\frac{S}{B(\t)P}\bigg)^{-\frac{2\g}{\s^2}+1}
    V_C\bigg(\frac{(B(\t)P)^2}{S},KP,\t\bigg)
\end{eqnarray*}
The result quoted in the literature corresponds to the case
where prices are expressed in the currency. If we consider a constant
barrier (setting $\g=r$), the above equation collapses to
\[
V_{DOC}(S,P,K,B,\g,\t)=
V_C(S,KP,\t)
-\bigg(\frac{S}{B} \bigg)^{-\frac{2r}{\s^2}+1}
V_C\bigg(\frac{B^2}{S},KP,\t\bigg)
\]
As it should. 

\subsection{Rebates}

Often, barrier options specify a rebate, an amount of money
paid to the holder if the barrier is hit in the case of a
knock-out option or not hit in the case of a knock-in option.
The premium that has to be paid for this provision can be
calculated in terms of a rebate option. For example, let us
consider a (generalized) down-and-out rebate option
$V_{DOR}(S,P,B,K,\g_1,\g_2,\t)$. This option pays the holder
$Ke^{\g_2\t}P$ at the first moment $\t>0$ for which
$S=Be^{\g_1\t}P$. The usual choice is $\g_1=\g_2=r$,
corresponding to a fixed barrier $B$ and a fixed rebate
$K$ in terms of money. The option can be characterized
by the following boundary conditions
\begin{gather*}
V_{DOR}(Be^{\g_1\t}P,P,B,K,\g_1,\g_2,\t) = Ke^{\g_2\t}P
\quad\mathrm{\mbox{for all }} \t>0 \\
V_{DOR}(S,P,B,K,\g_1,\g_2,0) = 0
\quad\mathrm{\mbox{for all }} S>BP
\end{gather*}
To solve this problem, we first construct a tradable
which has the proper value at the boundary
\[
\l R_\a(S,BP,\t) = Ke^{\g_2\t}P
\quad\mathrm{\mbox{for }} S=Be^{\g_1\t}P
\]
We find that this equation has two solutions, given by
\[
\l=\frac{K}{B}, \hspace{5mm} 
\a_{\pm} = \frac{a\pm b}{\s^2}, \hspace{5mm}
a=\g_1-\frac{1}{2}\s^2, \hspace{5mm} 
b=\sqrt{a^2+2\g_2\s^2}
\]
We see that $V_{DOR}-\l R_{\a_-}$ should vanish on the boundary.
Now we can apply essentially the same techniques as we did
for the down-and-out call to find that the rebate option
can be written as
\[ \begin{split}
V_{DOR}(S,P,B,K,\g_1,\g_2,\t) &= \frac{K}{B} \bigg(
  R_{\a_-}(S,BP,\t) \\
&+ V_D\big(R_{\a_+}(S,BP,\t),R_{\a_-}(S,BP,\t),
   \left(\frac{2b}{\s^2}\right)^2\t\big) \\
&- V_D\big(R_{\a_-}(S,BP,\t),R_{\a_+}(S,BP,\t),
   \left(\frac{2b}{\s^2}\right)^2\t\big)
\bigg) \end{split}
\]
Here $V_D(S,P,\t)$ is an asset-or-nothing digital option, which is
defined as
\[
V_D(S,P,\t) = S \Phi\left(\frac{\log\left(\frac{S}{P}\right)
+\frac{1}{2}\s^2\t}{\s\sqrt{\t}}\right)
\]
The up-and-out rebate can be found by using a generalized put-call
transformation. It is given by
\[
V_{UOR}(S,P,B,K,\g_1,\g_2,\t)=
V_{DOR}\bigg(P,S,\frac{1}{B},\frac{K}{B},-\g_1,-\g_1+\g_2,\t\bigg)
\]
What about knock-in rebates? We cannot simply use in-out parity to
find values for these options because of the difference in timing of
the payoff. Knock-in rebates are always paid at maturity, while
knock-out rebates are generally paid before maturity. However, we can
easily calculate the value of a knock-out rebate which does pay out at
maturity, simply by setting $\g_2=0$. This allows us to use in-out
parity after all and write
\begin{eqnarray*}
  V_{DIR}(S,P,B,K,\g_1,\t) 
  &=& KP-V_{DOR}(S,P,B,K,\g_1,0,\t) \\
  V_{UIR}(S,P,B,K,\g_1,\t) 
  &=& KP-V_{UOR}(S,P,B,K,\g_1,0,\t)
\end{eqnarray*}

\subsection{Double barriers}

Next, let us consider a call option with strike $K$ which knocks out
on two boundaries. The upper boundary is defined by $S=H e^{\g_1\t}
P$, the lower by $S=L e^{\g_2\t} P$. Again, the choice $\g_1=\g_2=r$
corresponds to constant boundaries.  We assume that $L<K<H$. A price
for such a call can be constructed in an iterative way. As a first
approximation, we consider an up-and-out call, knocking out on the
upper boundary
\[
V_{OC}(S,P,K,H,L,\g_1,\g_2,\t) \sim V_{UOC}(S,P,K,H,\g_1,\t)
\]
Of course, this overestimates the price, since this up-and-out call
has a positive value on the lower boundary. To compensate, we add a
correction term $V_1$ which has the opposite value on this boundary,
and which has value zero at maturity if $S>LP$.  In the same way as
before, we find that this correction term can be written as
\begin{eqnarray*}
  V_1
  &=& \left(\frac{H}{L}\right)^{\frac{2\g_1}{\s^2}+1}V_{UOC}
  \left(
    X_1,\frac{Y_1}{H},\frac{KL}{H},L,\g_1,\t
  \right) \\
  X_1
  &=& R_{\frac{2(\g_1-\g_2)}{\s^2}+1}(S,LP,\t)
  ,\quad
  Y_1=R_{\frac{2(\g_1-\g_2)  }{\s^2}}(S,LP,\t)
\end{eqnarray*}
But now, the value of the option is not vanishing on the upper
boundary. So we need another correction term $V_{-1}$, which
compensates this, and has value zero at maturity if $S<HP$.  Such a
term can be constructed as
\begin{eqnarray*}
  V_{-1}
  &=& \left(\frac{H}{L}\right)^{-\frac{2\g_1}{\s^2}-1}V_{UOC}
  \left(
    X_{-1},\frac{Y_{-1}}{H},\frac{KH}{L},\frac{H^2}{L},\g_1,\t
  \right) \\
  X_{-1}
  &=& R_{-\frac{2(\g_1-\g_2)}{\s^2}+1}\bigg(S,\frac{H^2}{L}P,\t\bigg) 
  ,\quad
  Y_{-1}=R_{ -\frac{2(\g_1-\g_2)}{\s^2}}\bigg(S,\frac{H^2}{L}P,\t\bigg)
\end{eqnarray*}
Continuing in this way, we find an infinite series representation for
the price of the out-call. The successive terms in this sum represent
up-and-out calls which are more and more out of the money, so that the
corrections terms rapidly become smaller, and in practice we need only
a few terms to find a price with reasonable accuracy. The final
formula becomes
\[
V_{OC}(S,P,K,H,L,\g_1,\g_2,\t) = \sum_{n=-\infty}^{\infty} V_n
\]
where
\begin{eqnarray*}
  V_n
  &=&\left(\frac{H}{L}\right)^{n\left(\frac{2\g_1}{\s^2}+1\right)}V_{UOC}
  \left(
    X_n,\frac{Y_n}{H},\frac{KL^n}{H^n},\frac{HL^n}{H^n},\g_1,\t
  \right) \\
  X_n
  &=&R_{\frac{2n(\g_1-\g_2)}{\s^2}+1}\bigg(S,\frac{HL^n}{H^n}P,\t\bigg)
  ,\quad
  Y_n=R_{\frac{2n(\g_1-\g_2)}{\s^2}  }\bigg(S,\frac{HL^n}{H^n}P,\t\bigg)
\end{eqnarray*}
The functions $V_n$ have the property that $V_n+V_{-n}$ vanishes on
the upper boundary and $V_n+V_{-n+1}$ vanishes on the lower boundary.
Furthermore, $V_n=0$ at maturity if $L<S<H$ and $n\neq 0$.  We have
checked this result against that of \cite{KunitomoIkeda92} and found
identical results and rates of convergence. By using in-out parity, we
find the following formula for the price of an knock-in call with two
barriers
\[
V_{IC}(S,P,K,H,L,\g_1,\g_2,\t) = V_{UIC}(S,P,K,H,\g_1,\t)-
\sum_{n\neq 0} V_n
\]
Generalized put-call transformations give the price for the
corresponding double boundary put options
\begin{eqnarray*}
  V_{OP}(S,P,K,H,L,\g_1,\g_2,\t)
  &=& KV_{OC}\bigg(P,S,\frac{1}{K},
  \frac{1}{L},\frac{1}{H},-\g_2,-\g_1,\t\bigg) \\
  V_{IP}(S,P,K,H,L,\g_1,\g_2,\t)
  &=& KV_{IC}\bigg(P,S,\frac{1}{K},
  \frac{1}{L},\frac{1}{H},-\g_2,-\g_1,\t\bigg)
\end{eqnarray*}
What will happen when $K<L<H$ in the case of a double
barrier call option? In the derivation of the above formulae
we have made essential use of the fact that $V_{UOC}(S,P,K,H,\g_1,0)$
vanishes for $S<LP$. However, this is no longer true when $K<L$. 
So in order to find a price for this configuration of barriers,
we need a new building block, the 'left-clipped' up-and-out call.
It can be defined as follows
\[ 
\begin{split}
  V_{UOC}^c(S,P,K,H,L,\g,\t)
  &=V_C^{H,L}(S,KP,\t)\\
  &-V_C^{H,L}\bigg(R_\a(S,HP),
  \frac{K}{H}R_{\a+1}(S,HP),\t\bigg)
\end{split}
\]
where the option $V_C^{H,L}$ is defined by the doubly clipped payoff
\[
V_C^{H,L}(S,KP,0)=\left\{\begin{array}{ll}
    S-KP & \mbox{for $LP<S<HP$}\\
    0 & \mbox{otherwise} \end{array} \right.
\]
Its price is thus given by $V_C^{H,L}=V_C^{L+}-V_C^{H+}$.  Now
$V_{UOC}^c(S,P,K,H,L,\g_1,0)$ does vanish for $S<LP$.  Therefore in
the case that $K<L<H$ we need to replace the function $V_n$ by the
following modified functions, a fact which was not recognized in
\cite{KunitomoIkeda92}:
\[
\tV_n
=\left(\frac{H}{L}\right)^{n\left(\frac{2\g_1}{\s^2}+1\right)}
V_{UOC}^c\left(X_n,\frac{Y_n}{H},\frac{L^n}{H^n}K,\frac{L^n}{H^n}H,
  \frac{L^n}{H^n}L,\g_1,\t \right)
\]
As a final note, let us mention that continuous dividend payments with
rate $q$ can easily be incorporated by making the usual substitution
$S\rightarrow Se^{-q\t}$. Simultaneously, all $\g$'s that appear in
the definition of boundaries should be adjusted like
$\g\rightarrow\g-q$.

\subsection{Lookback options}

We now turn to lookback options. Let us consider a floating strike
lookback put. It pays the owner the difference between the maximum
realized price and the spot price of some asset at expiry. This
maximum is usually defined with respect to a given currency. Assuming
that we are in a deterministic bond setting, it can be written as
\[
S_{\max} = \max_{0\leq t\leq T} \frac{S(t)}{e^{\g\t}P(t,T)}
\]
with $\g=r$. We will consider a slightly more general definition,
leaving $\g$ arbitrary. Interestingly, this option can also be
described in terms of a boundary problem
(Ref.~\cite{GoldmanSosinGatto79}).  If we introduce $J=S_{\max}P$ and
denote the price of the option by $V_{LP}(S,J,\g,\t)$, the boundary
conditions are
\begin{eqnarray*}
  V_{LP}(S,J,\g,0) &=& J-S
  ,\quad 
  J\geq S \\
  \left. \pd_J V_{LP}(S,J,\g,\t) \right|_{S=e^{\g\t}J}=0
  ,\quad\mathrm{\mbox{for all }} \t
\end{eqnarray*}
The latter condition allows to roll-up the position self-financingly
when $S$ reaches a new high. It implies that at the boundary
$S=e^{\g\t}J$ all money is invested in the stock. Now let us try a
solution of the following form
\[
V_{LP}(S,J,\g,\t) 
= V_P(S,J,\t)+V_P(\l R_\a(S,J,\t),\m R_\b(S,J,\t),(\a-\b)^2\t)
\]
where $V_P(S,J,t)$ is the price of a plain vanilla put with strike
$S_{\max}$. One can check that this is indeed a solution, provided
that we have
\[
\l=\m=\frac{1}{k}
,\quad
\a=1-k
,\quad
b=1
,\quad
k=\frac{2\g}{\s^2}
\]
Note that we must have $\g>0$ or else the price blows up. In more
detail, the solution is
\begin{eqnarray*}
  \lefteqn{V_{LP}(S,J,\g,\t)=}\\
  &=&J\Phi(-d_2)-S\Phi(-d_1)+\frac{S}{k}\left(\Phi(d_1)
    -e^{-\g\t}\left(\frac{e^{\g\t} J}{S}\right)^k
    \Phi(d_1-k\S)\right) 
\end{eqnarray*}
where
\[
d_{1,2}=\frac{\log\left(\frac{S}{J}\right)\pm\frac{1}{2}\S^2}{\S}
,\quad
\S=\s\sqrt{\t}
\]
From this, we find the following delta's
\begin{eqnarray*}
  \pd_S V_{LP} &=&-\Phi(-d_1)+k^{-1}\Phi(d_1)+
  (1-k^{-1})e^{-\g\t}\left(\frac{e^{\g\t} J}{S}\right)^k\Phi(d_1-k\S) \\
  \pd_J V_{LP} &=& 
  \Phi(-d_2)-\left(\frac{e^{\g\t}J}{S}\right)^{k-1}\Phi(d_1-k\S)
\end{eqnarray*}
In a very similar way, we can derive the value of the floating strike
lookback call option. If we define $J=S_{\min}P$, its value is given
by
\[
V_{LC}(S,J,\g,\t) = V_C(S,J,\t)+\frac{1}{k}
V_C(R_{1-k}(S,J,\t),S,k^2\t)
\]

\section{Discussion and outlook}
\label{sec:discussion-outlook}

We have shown in this paper that the formalism put forward in
Ref.~\cite{HooglandNeumann99a} provides a powerful framework for the
pricing of path-dependent contingent claim pricing. The formulation of
the pricing problem in terms of tradables leads to more transparent
formulae with, by construction, clear financial interpretations. The
scaling symmetry which should be satisfied at any time by the claim
prices provides a very powerful check when doing computations.
Exploiting symmetries of the governing PDE leads to large families of
related claims. Also put-call symmetries follow naturally in the
formalism.

In the following paper~\cite{HooglandNeumann99c} we will show that the
pricing of claims can be formulated in a geometric way. This provides
an alternative and more fundamental formulation for the pricing of
contingent claims than the one using semi-martingales. It also
provides a more intuitive formulation by focussing on the symmetries
of the prices of tradables.  

By formulating the pricing of claims in terms of tradables, we can
also clarify and extend results on American-type options and stochastic
volatility models. This will be discussed in future papers.

\bibliographystyle{alpha} 
\bibliography{relativity}

\end{document}